\documentclass[10pt,conference]{IEEEtran}

\usepackage{cite}
\usepackage{listings}
\newcommand{\code}[1]{\texttt{#1}}

\usepackage{xcolor}

\definecolor{codegreen}{rgb}{0,0.6,0}
\definecolor{codepurple}{rgb}{0.58,0,0.82}
\definecolor{codegray}{rgb}{0.5,0.5,0.5}
\definecolor{boxcolour}{rgb}{0.5,0.5,0.5}
\definecolor{backcolour}{rgb}{1,1,1}

\lstdefinestyle{screencap}{
    numbers=none,
    numbersep=2pt,
    basicstyle=\ttfamily\scriptsize,
    breakatwhitespace=false,
    breaklines=false,
    captionpos=b,
    keepspaces=true,
    showspaces=false,
    showstringspaces=false,
    showtabs=false,
    frame=single
}

\lstdefinestyle{scripts}{
    numberstyle=\tiny\color{codegray},
    numbers=none,
    numbersep=2pt,
    basicstyle=\ttfamily\scriptsize,
    breakatwhitespace=false,
    breaklines=true,
    captionpos=b,
    keepspaces=true,
    showspaces=false,
    showstringspaces=false,
    showtabs=false,
    tabsize=2,
    language=bash,
    frame=single,
    rulecolor=\color{boxcolour}
}

\usepackage{amsmath,amssymb,amsfonts}
\usepackage{algorithmic}
\usepackage{textcomp}
\usepackage{xcolor}
\def\BibTeX{{\rm B\kern-.05em{\sc i\kern-.025em b}\kern-.08em
    T\kern-.1667em\lower.7ex\hbox{E}\kern-.125emX}}
\usepackage{cite}
\usepackage[pdftex]{graphicx}
\DeclareGraphicsExtensions{.pdf,.jpeg,.png}
\usepackage{url}
\usepackage{tikz}
\usetikzlibrary{shapes.multipart}
\usetikzlibrary{shapes.geometric}
\usetikzlibrary{positioning}
\usepackage[acronym]{glossaries}

\newacronym{hpc}{HPC}{high performance computing}
\newacronym{wlm}{WLM}{workload manager}
\newacronym{qec}{QEC}{quantum error correction}
\newacronym{qem}{QEM}{quantum error mitigation}
\newacronym{nisq}{NISQ}{noisy, intermediate-scale quantum} 
\newacronym{mpi}{MPI}{message passing interface}
\newacronym{api}{API}{application programming interface}
\newacronym{rpc}{RPC}{remote procedure call}
\newacronym{hhl}{HHL}{Harrow-Hassidim-Lloyd}
\newacronym{qpu}{QPU}{quantum processing unit}
\newacronym{cpu}{CPU}{central processing unit}
\newacronym{ghz}{GHZ}{Greenberger-Horne-Zeilinger}
\newacronym{mpmd}{MPMD}{multiple program, multiple data}
\newacronym{qasm}{QASM}{quantum assembly language}
\newacronym{isc}{ISC}{International Supercomputing Conference}
\newacronym{mtbf}{MTBF}{mean time between failures}

\usepackage[english]{babel}
\usepackage[utf8x]{inputenc}

\def\BibTeX{{\rm B\kern-.05em{\sc i\kern-.025em b}\kern-.08em
    T\kern-.1667em\lower.7ex\hbox{E}\kern-.125emX}}

\begin{document}

\title{A Hybrid Classical-Quantum HPC Workload}

\author{\IEEEauthorblockN{Aniello Esposito\IEEEauthorrefmark{1},
Jessica R. Jones\IEEEauthorrefmark{2}, Sebastien Cabaniols\IEEEauthorrefmark{3} and
David Brayford\IEEEauthorrefmark{4}}
\IEEEauthorblockA{\textit{HPC/AI EMEA Research Lab}\\
\textit{Hewlett Packard Labs} \\
Location: \IEEEauthorrefmark{1}Basel Switzerland,
\IEEEauthorrefmark{2}Bristol UK,
\IEEEauthorrefmark{3}Grenoble France,
\IEEEauthorrefmark{4}Munich Germany \\
Email: \IEEEauthorrefmark{1}aniello.esposito@hpe.com,
\IEEEauthorrefmark{2}j.r.jones@hpe.com,
\IEEEauthorrefmark{3}sebastien.cabaniols@hpe.com,
\IEEEauthorrefmark{4}david.kenneth.brayford@hpe.com}}
%
%
%
%
\maketitle
%
\begin{abstract}
A strategy for the orchestration of hybrid classical-quantum workloads on supercomputers featuring quantum devices is proposed. The method makes use of heterogeneous job launches with Slurm to interleave classical and quantum computation, thereby reducing idle time of the quantum components. To better understand the possible shortcomings and bottlenecks of
such a workload, an example application is investigated that offloads parts of the computation to a quantum device.  It executes 
on a classical \gls{hpc} system, with a server mimicking the quantum device, within the \gls{mpmd} paradigm in Slurm. 
Quantum circuits are synthesized by means of the \texttt{Classiq} software
suite according to the needs of the scientific application, and the Qiskit Aer circuit simulator computes the 
state vectors. The \gls{hhl} quantum algorithm for linear systems of equations is used to solve the algebraic problem from the discretization of a linear differential equation. 
Communication takes place over the \gls{mpi}, which is broadly employed in the \gls{hpc} community. Extraction of state vectors and circuit synthesis are the most time consuming, while communication is negligible in this setup. 
The present test bed serves as a basis for more advanced hybrid workloads eventually involving a real quantum device. 
%
\end{abstract}
\section{Introduction}
\IEEEPARstart{Q}{uantum} computers hold the potential to solve certain very difficult problems with moderate input sizes efficiently but accessibility and usability lag behind. On the other hand, supercomputers and their data-intensive applications are large but operate with proven tools developed over decades.
Several promising algorithms for quantum architectures have been developed over the past decades~\cite{montanaro2016quantum}. However, as was explained in \cite[Sec. 1.1]{davenport2023practical}, parallelism is not as easy to realise in quantum algorithms. A more suitable approach therefore, given the small size of early \gls{nisq} devices, is to offload portions of a classical code that would most benefit from quantum speed-up.
Many \gls{hpc} applications could profit from this approach in a hybrid workload, but how that might happen in a practical sense has not been standardized yet. It is therefore crucial that today's computational scientists are able to adapt early on to make use of these machines as they become available, while ideally keeping as much as possible from the proven \gls{hpc} ecosystem, in order to shape hybrid workflows to their needs. In \cite{davenport2023practical} some of the challenges that will need to be addressed to make a hybrid classical-quantum supercomputer truly useful to the wider scientific community are explained.
%
Assuming that those challenges can be addressed, a hybrid classical-quantum system can be thought of as broadly similar to any other \gls{hpc} system with more than a single node architecture, e.g. with a mixture of accelerators such as GPUs and FPGAs. That means that most of the tools at our disposal today, such as Slurm~\cite{slurm} and \gls{mpi}~\cite{mpi40standard}, can be repurposed. Slurm already supports heterogeneous jobs and can be configured to schedule hybrid classical-quantum jobs on a suitable hybrid machine.

The first part of this work presents an example hybrid workload as described above. It makes use of the \gls{hhl} quantum algorithm to solve a system of linear equations repeatedly, where 
a quantum device is mimicked by a circuit simulator within the \gls{mpmd} paradigm. This testbed allows the investigation of possible shortcomings and bottlenecks of the workload design, which is the main purpose and contribution of this part. A solid understanding of this preliminary stage is crucial before including a real quantum device, which would then allow investigation of the performance of both the hardware and software. In the second part, a general and more sophisticated strategy for hybrid classical-quantum workloads is proposed that employs heterogeneous Slurm jobs. This approach allows the idle time of a quantum device to be further reduced, thereby increasing the efficiency of hybrid workloads. This is an improvement over \gls{mpmd} from the first part, where the quantum device is not released during a Slurm job and can be left idle.
%
%
%
\section{The Hybrid Workflow}
\subsection{General Considerations}
\begin{figure}[!t]
\centering
\includegraphics[width=\columnwidth]{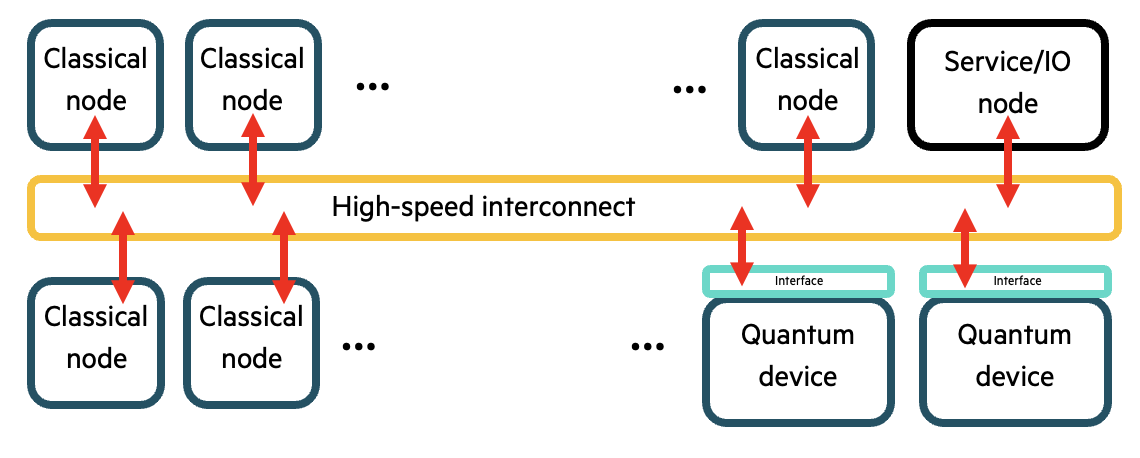}
\caption{Schematic representation of a tightly coupled hybrid system, where quantum devices access via an interface the same high-speed network as the classical compute nodes and service nodes.}
\label{fig:hybrid_scheme}
\end{figure}

%
A realistic architecture of a hybrid classical-quantum system in the near future could consist of a large number of classical compute nodes, as found in today's supercomputers, and at least one or two orders of magnitude fewer nodes with quantum hardware, likely \gls{nisq} devices with 10-100 qubits each, sharing a common high-speed interconnect as shown in Fig.~\ref{fig:hybrid_scheme}.
The quantum devices are exposed as regular compute nodes to a \gls{wlm} using a hardware interface that
is assumed to be compatible with the possibly very different cooling requirements of the various architectures. 
%
%
From an algorithmic point of view, offloading portions of a classical code to quantum devices presents another challenge, namely that of data transfer speed, which will affect how those portions are selected. Hoefler et.al.~\cite{hoefler2023disentangling} demonstrated, with some fairly pessimistic, if realistic, assumptions about I/O speed, that it is necessary to minimise the amount of data being transferred to benefit from the advantages conferred by quantum computation.  Algorithm selection is therefore key in realising quantum advantage.
The problem of \gls{qec} is ignored for the moment.
However, it is important to note that this must be addressed before widespread adoption of systems of this type in the \gls{hpc} community.
See \cite[Sec. 1.2]{davenport2023practical} for further discussion on this issue.
%
%
%
%
%
%
%
%
%
\subsection{Example Application}\label{sec:setup}
\begin{figure}[!t]
\centering
\includegraphics[width=0.9\columnwidth]{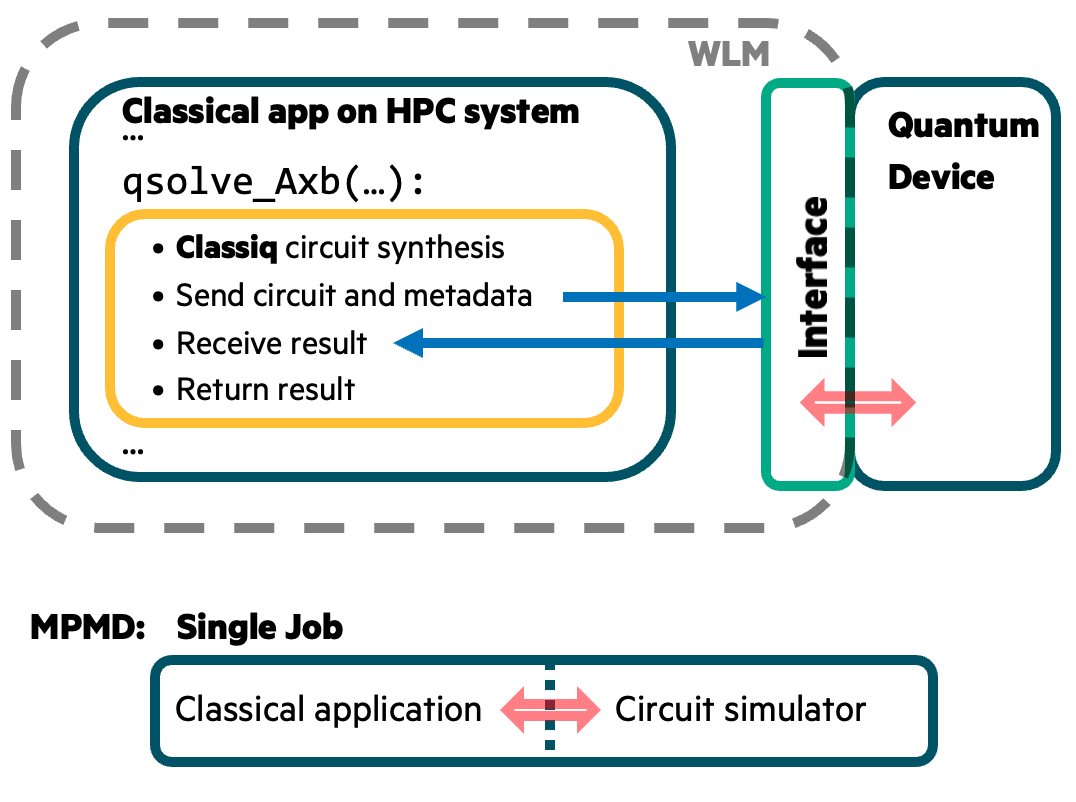}
\caption{Schematic representation of the interaction between a classical application running on the \gls{hpc} system and an interface server for a quantum device. A \gls{wlm} spans a single job that includes both components running in a MPMD model. The subroutine \texttt{qsolve\_Axb(\ldots)} handles the synthesis of the quantum circuit depending on $A$ and $b$ on the \gls{hpc} system, as well as the communication with the quantum device. For the present case, a quantum circuit simulator mimics the quantum device.}
\label{fig:sim1}
\end{figure}

\begin{figure}[!t]
\centering
\includegraphics[width=0.9\columnwidth]{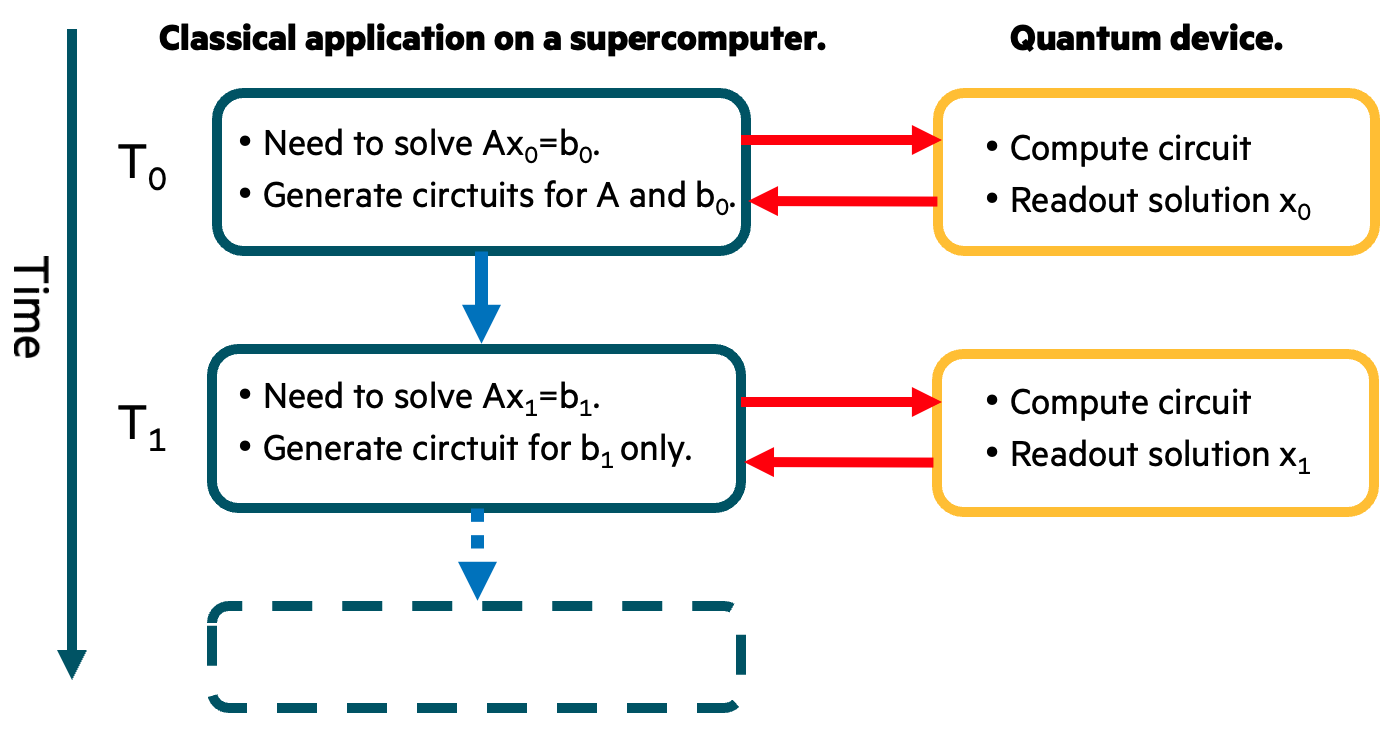}
\caption{Iterative solution of a linear system of equations $Ax=b$ as it is used for the time evolution 
of a discretized partial differential equation. At each time step the circuit for the \gls{hhl} algorithm is synthesized for a different right hand 
side and the solution is computed on a quantum device (quantum circuit simulator in the present case). Note that the matrix $A$ is synthesised only once as long as the 
discretization grid does not change.}
\label{fig:sim2}
\end{figure}

The discretization of linear differential equations is a standard numerical method for solving problems in engineering. The resulting algebraic problem in the form of a system of linear equations $Ax=b$ can be solved in principle on a quantum device by the \gls{hhl}\cite{harrow2009quantum} algorithm.
Fig~\ref{fig:sim1} shows a classical application calling a subroutine \code{qsolve\_Axb(\ldots)} that passes the the matrix $A$ and right-hand-side $b$ and returns the solution
$x$. 
The routine 
synthesizes the \gls{hhl} quantum circuits based on $A$ and $b$ by means of the \texttt{Classiq}~\cite{classiq} software
suite. The circuits are represented by strings of quantum circuit intermediate representation code (\gls{qasm})\cite{bishop2017qasm}, and these are transferred via \code{MPI\_Send/Recv} calls using mpi4py\cite{dalcin2021mpi4py}. \gls{mpi} has been 
chosen for communicating data because of its established role in supercomputing, but other mechanisms are also 
imaginable, such as the Maestro middleware\cite{haine2021middleware}.  
The quantum device that is supposed to process the quantum circuit is accessed through an interface 
running in another application. Though, in the present case, the quantum device is
mimicked by the \texttt{Qiskit Aer}~\cite{qiskit-aer} circuit simulator, which computes the final state vector and 
sends the solution back to the classical application.
In the simplest scenario, both applications are executed within a single job using the \gls{mpmd} model, where the job is allocated and started by Slurm (see appendix~\ref{sec:appendix}). The classical application and the quantum circuit simulator share a single \code{MPI\_COMM\_WORLD}, which required minor changes in
\texttt{Qiskit Aer}. Alternatively, dynamic process management in \gls{mpi} could be used to create an inter-communicator from the individual communicators of the two applications in order to send and receive information.
In the case of a linear time-dependent differential equation, a system of linear equations needs to be repeatedly solved, as illustrated in Fig~\ref{fig:sim2}.
When a discretization grid does not change over time, the matrix $A$ remains the same. The \texttt{Classiq}
software allows the right-hand-side $b$ to be synthesized separately, and then combined the final circuit without 
having to synthesize $A$ again. In the present use case, the circuit for every time step is sent to the quantum device, which then computes the solution and sends it back to the classical application where the $b$ vector is assembled for the next time step. Experiments have been conducted on a HPE Cray EX system featuring two AMD EPYC 7763 (Milan) CPUs per node, giving a total of 128 cores and 512GB memory connected via the Slingshot interconnect. The NumPy~\cite{harris2020array}
and mpi4py packages made use of the highly tuned math and \gls{mpi} libraries from the HPE Cray programming environment. Only moderate matrices and right-hand-sides of sizes up to $N=64$ have been considered and the number of qubits needed for the quantum algorithm is $O(\log(N))$. The present implementation uses Python and data is represented by NumPy arrays, but an implementation in 
C/C++ that passes pointers or references to arrays of floating point numbers is straightforward.
This workload has been demonstrated during the \gls{isc} 2023.
\subsection{Results and Discussion}\label{sec:results}
The most time consuming portion of the workload in Fig.~\ref{fig:sim2} is the synthesis of
$A$ and $b$ into quantum circuits, followed by the extraction of the state vector from the circuit simulator. 
The former is most probably due to the connection of the \texttt{Classiq} software to an 
external server, which could be highly improved by using an on-premise solution, and further optimizations
of the synthesis can be considered once this bottleneck has been removed. The matrix $A$ is usually sparse in terms of non-zero entries depending on the discretization scheme, but for the circuit generation an efficient decomposition in tensor products of Pauli matrices is more important. The full extraction of the state vector is not necessary 
as long as only a portion of the simulation domain in real-space is of interest, although this could be
insufficient in a time-dependent problem. The precision of the solution is another peculiarity of the quantum 
algorithm that deviates by several percent from the classical solution. As long as precision can be traded
for efficiency this is not an issue, but otherwise one has to include more qubits for precision and error
correction. Another approach would be use the approximate solution as an input to an iterative refinement 
procedure. Finally, the complexities of the individual portions of the quantum algorithm are summarized in 
Table~\ref{tab:complexity}.
\begin{table}[!t]
\caption{Complexity of the various phases of solving a linear system of equations in the hybrid workflow and a comparison to the classical counterpart. The synthesis of the circuit for the matrix $A$ has to be done only once, depending on the change in discretization, and $\kappa$ is the condition number of $A$.} 
\label{tab:complexity}
\centering
\begin{tabular}{|c|c|c|}
\hline
operation & quantum & classical\\
\hline
Solving $Ax=b$               & $O(\kappa \log(N))$ & O(N)\\
Synthesis of circuit for $b$ & $O(\log(N)$ & N/A \\
Synthesis of circuit for $A$ & $O(\log(N)$ & N/A \\
Readout information from quantum device & $O(1)-O(N)$ & N/A \\
\hline
\end{tabular}
\end{table}
Assuming that only part of the state vector is needed and that the injection of the circuit in the quantum device, as well as the circuit transfer time, remain negligible, 
the conditions under which this hybrid workload can eventually outperform its purely classical counterpart are given.
\subsection{Anatomy of an Improved Hybrid Workload}\label{sec:anatomy}
\begin{figure}[!t]
\centering
\includegraphics[width=\columnwidth]{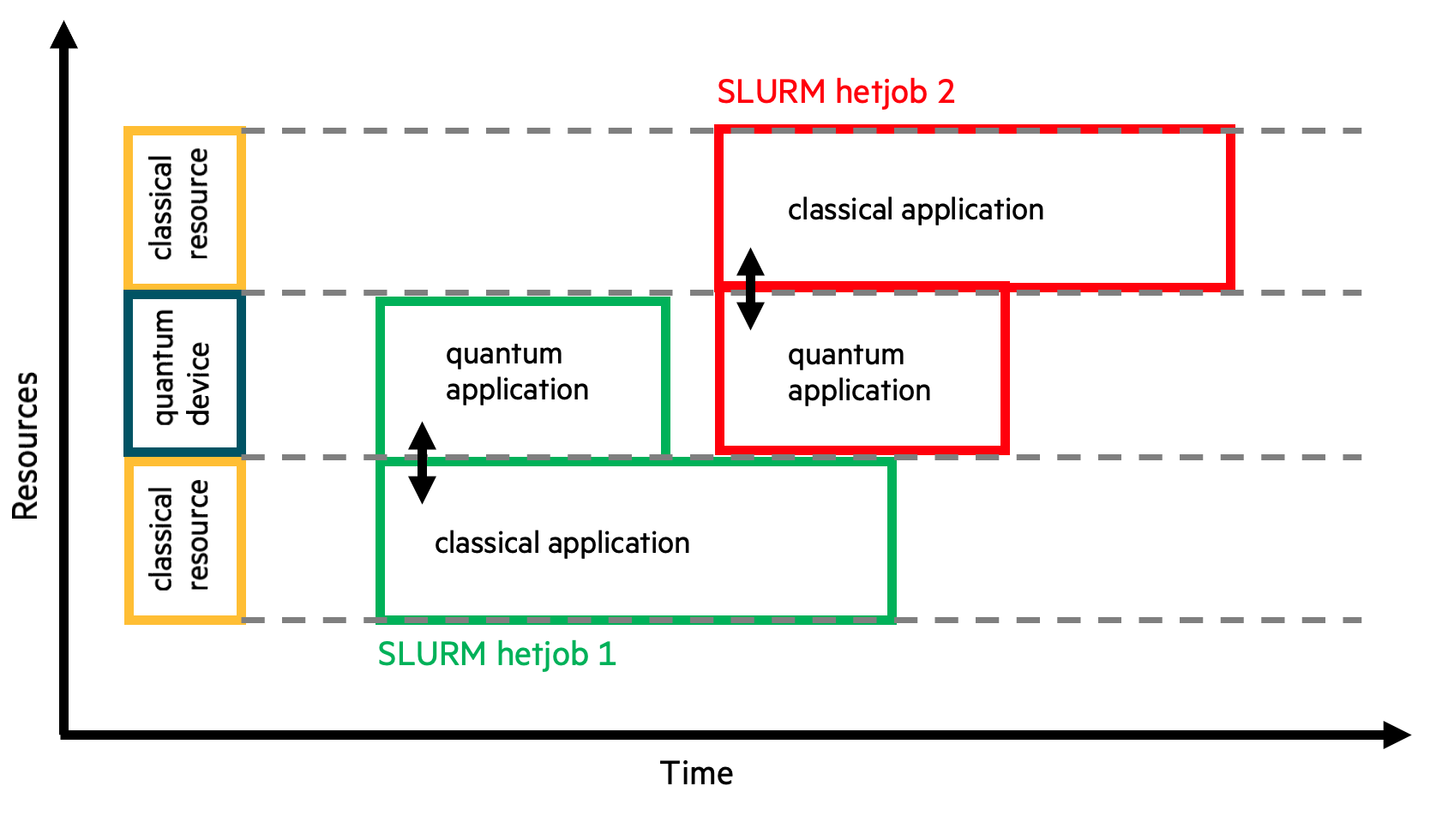}
\caption{Schematic representation of two Slurm heterogeneous jobs requiring a quantum device that is exposed as a compute node. As soon as the quantum device is no longer needed by the first heterogeneous job it can be released while the classical part continues to run. The second heterogeneous job can start using the quantum device.}
\label{fig:sim3}
\end{figure}

The \gls{mpmd} model used in Sec.~\ref{sec:setup} is simple, but it blocks a quantum device for the whole duration 
of the classical application and so potentially wastes precious resources. A reduction of this idle time can be achieved by using the Slurm support for heterogeneous jobs (\code{hetjobs}) to split a job across differing hardware.
%
%
%
A simple scenario consists of two heterogeneous jobs \code{\{job1,job2\}} , each requiring classical and quantum computing resources. 
As is typical in \gls{hpc}, the two jobs are submitted to a queue.  Once resources are available to start, both the classical and quantum parts of \code{job1} begin.  At a crucial point in the execution, there is a synchronisation to allow the classical part to wait on results from its quantum counterpart. The classical part polls the quantum resource for completion.  Once the quantum computation is finished, the resource is freed and then immediately consumed by the quantum part of \code{job2}, which has been waiting in the queue for the resource to become available.  This is illustrated in Fig~\ref{fig:sim3}, where it is assumed that the quantum device is the bottleneck, since their availability will initially be limited compared to that of traditional hardware.
Sources for the job scripts can be found in the appendix~\ref{sec:appendix}. Usually, the consumption of classical and quantum resources does not start at the same time, however this can be achieved by splitting, checkpointing, and restarting the classical part appropriately.  
Here follow screen captures of the output of \code{squeue} at critical moments during the execution of the two \code{hetjobs}. The various states of a job can be ``pending'' (\code{PD}), ``completing'' (\code{CG}), or ``running'' (\code{R}).
\begin{lstlisting}[style=screencap, caption={job2 waits while job1 uses the quantum resource}, label={code:pendingQ1}]
JOBID PARTITION  NAME   USER ST  TIME NODES NODELIST(REASON)
105756+1      Q   job2  seb PD  0:00    1      (Resources)
105754+1      Q   job1  seb  R  0:21    1           qnode1
105756+0    HPC   job2  seb PD  0:00    2      (Resources)
105754+0    HPC   job1  seb  R  0:21    2    hpcn[136-137]
\end{lstlisting}
In Listing~\ref{code:pendingQ1}, \code{job1} has been allocated the quantum resource, \code{qnode1} and, the two \gls{hpc} machines are waiting for the end of the quantum computation. \code{job2} is waiting on resources because the quantum machine, \code{qnode1} is busy.
\begin{lstlisting}[style=screencap, caption={The quantum part of \code{job1} is completing.}, label={code:completingQ1}]
JOBID PARTITION  NAME  USER ST  TIME NODES NODELIST(REASON)
105754+1     Q    job1 seb CG  0:31    1          qnode1
105756+1     Q    job2 seb PD  0:00    1      (Resources)
105756+0   HPC    job2 seb PD  0:00    2      (Resources)
105754+0   HPC    job1 seb  R  0:33    2    hpcn[136-137]
\end{lstlisting}
In Listing~\ref{code:completingQ1}, the quantum part of \code{job1} is done; the quantum resource is being liberated, while the classical part of \code{job1} is still running.
Meanwhile, \code{job2} is still pending as it waits for the resources it needs to become available.
\begin{lstlisting}[style=screencap, caption={Classical part of \code{job1} is still computing, while \code{job2} is still pending, but the Q resource is almost ready again now.}, label={code:runningClassic}]
JOBID PARTITION  NAME  USER ST  TIME NODES NODELIST(REASON)
105756+1    Q     job2 seb PD  0:00    1      (Resources)
105756+0  HPC     job2 seb PD  0:00    2      (Resources)
105754+0  HPC     job1 seb  R  0:45    2    hpcn[136-137]
\end{lstlisting}
In Listing~\ref{code:runningClassic} \code{job1}'s classical part is still computing, while 
\code{job2} is still pending, but the quantum resource is almost ready again now.
\begin{lstlisting}[style=screencap, caption={Classical part of \code{job1} is still computing, while \code{job2} is running}, label={code:runningQ2}]
JOBID PARTITION  NAME  USER ST  TIME NODES NODELIST(REASON)    
105756+1    Q     job2 seb  R  0:01    1          qnode1
105754+0  HPC     job1 seb  R  1:01    2   hpcn[136-137]
105756+0  HPC     job2 seb  R  0:01    2   hpcn[138-139]
\end{lstlisting}
Listing~\ref{code:runningQ2} shows \code{job1} is continuing its classical computation.
\code{job2} is now running, with the quantum computation in progress and the classical computation eventually waiting for the results (blocking \gls{mpi} call).
%
%
%
%
\section{Conclusion}\label{sec:conclusion}
A hybrid classical-quantum workload for the repeated solution of a system of linear equations has been executed 
on a single HPE Cray EX system within the \gls{mpmd} model of Slurm. The circuits are generated by a classical 
scientific application and then evaluated by a simulator imitating a quantum device. Synthesis of a circuit is ideally 
done locally and not remotely, which is really an administrative or legal (licensing) issue, rather than a technical one. 
The lower precision of the results from a quantum device needs to be taken into account and improved 
by either using more qubits and an improved algorithm, or by classical techniques such as iterative refinement
for this particular use case. Extraction of the full state vector has to be avoided if possible. 
For the more flexible case, where a quantum resource does not have to be bound to a classical application until
its termination in an \gls{mpmd} job, a strategy using heterogeneous jobs in Slurm has been presented, where 
the quantum device can be freed up during the job run-time, and used by the next heterogeneous job.
In the next step the use case will be executed as a heterogeneous job to profile the framework and eventually, the circuit simulator will be replaced by an actual quantum device.
%
\appendix\label{sec:appendix}
Listing~\ref{code:mpmd} shows the \gls{mpmd} Slurm job script for the example in Sec.~\ref{sec:setup}
\begin{lstlisting}[caption={Slurm MPMD job script},label={code:mpmd}]
#!/bin/bash

#SBATCH -N 1
#SBATCH -t 10
#SBATCH --exclusive

srun -n 2 -c 1  -u -l --multi-prog multi.conf


\end{lstlisting}
with the configuration file 
\begin{lstlisting}[caption={\code{multi.conf} Slurm MPMD configuration file},label={code:mpmd_conf}]
0 python hhl_demo.py
1 python circuit_simulator.py
\end{lstlisting}
The heterogeneous Slurm jobs shown in Sec.~\ref{sec:anatomy} are based on a script like the following
%
\begin{lstlisting}[caption={\code{hetjob.sh}, the main job script},label={code:hetjobsh}]
#!/bin/bash
#
# The first allocation is the classic HPC allocation and will survive
# the computational tasks even when the quantum part of the job is finished.
#
# The -C quantum is our quantic resource selector and we only have one (mimicked by
# requiring qnode1, a unique, particular system).

#SBATCH -N2
#SBATCH -C classic
#SBATCH -p HPC
#SBATCH --exclusive
#SBATCH hetjob
#SBATCH -N1
#SBATCH -C quantum
#SBATCH -p Q
#SBATCH -w qnode1
#SBATCH --exclusive

srun -n2 --ntasks-per-node=1 python hhl_demo.py : -n1 --ntasks-per-node=1 python circuit_simulator.py

\end{lstlisting}
%
%
%
%
\section*{Acknowledgements}
The authors would like to thank Alfio Lazzaro for proof reading and helpful comments, Frédéric Ciesielski and Yann Maupu for their Slurm expertise and testbed, and also the team 
at \texttt{Classiq} for allowing the authors access to their professional tool suite and for continuous support during 
our collaboration.  Thanks also are due to the reviewers, whose comments helped guide improvements to this work.
%
%
%
\bibliographystyle{IEEEtran}
\bibliography{refs}

\begin{thebibliography}{10}
\providecommand{\url}[1]{#1}
\csname url@samestyle\endcsname
\providecommand{\newblock}{\relax}
\providecommand{\bibinfo}[2]{#2}
\providecommand{\BIBentrySTDinterwordspacing}{\spaceskip=0pt\relax}
\providecommand{\BIBentryALTinterwordstretchfactor}{4}
\providecommand{\BIBentryALTinterwordspacing}{\spaceskip=\fontdimen2\font plus
\BIBentryALTinterwordstretchfactor\fontdimen3\font minus
  \fontdimen4\font\relax}
\providecommand{\BIBforeignlanguage}[2]{{%
\expandafter\ifx\csname l@#1\endcsname\relax
\typeout{** WARNING: IEEEtran.bst: No hyphenation pattern has been}%
\typeout{** loaded for the language `#1'. Using the pattern for}%
\typeout{** the default language instead.}%
\else
\language=\csname l@#1\endcsname
\fi
#2}}
\providecommand{\BIBdecl}{\relax}
\BIBdecl

\bibitem{montanaro2016quantum}
A.~Montanaro, ``Quantum algorithms: an overview,'' \emph{npj Quantum
  Information}, vol.~2, no.~1, pp. 1--8, 2016.

\bibitem{davenport2023practical}
J.~H. Davenport, J.~R. Jones, and M.~Thomason, ``A practical overview of
  quantum computing: Is exascale possible?'' 2023.

\bibitem{slurm}
\BIBentryALTinterwordspacing
``Slurm workload manager,'' \url{https://slurm.schedmd.com/}. [Online].
  Available: \url{https://slurm.schedmd.com/}
\BIBentrySTDinterwordspacing

\bibitem{mpi40standard}
\BIBentryALTinterwordspacing
{Message Passing Interface Forum}, \emph{{MPI}: A Message-Passing Interface
  Standard Version 4.0}, Jun 2021. [Online]. Available:
  \url{https://www.mpi-forum.org/docs/mpi-4.0/mpi40-report.pdf}
\BIBentrySTDinterwordspacing

\bibitem{hoefler2023disentangling}
\BIBentryALTinterwordspacing
T.~Hoefler, T.~H\"{a}ner, and M.~Troyer, ``Disentangling hype from
  practicality: On realistically achieving quantum advantage,''
  \emph{Communications of the ACM}, vol.~66, no.~5, p. 82–87, apr 2023.
  [Online]. Available: \url{https://doi.org/10.1145/3571725}
\BIBentrySTDinterwordspacing

\bibitem{harrow2009quantum}
A.~W. Harrow, A.~Hassidim, and S.~Lloyd, ``Quantum algorithm for linear systems
  of equations,'' \emph{Physical review letters}, vol. 103, no.~15, p. 150502,
  2009.

\bibitem{classiq}
\BIBentryALTinterwordspacing
``Classiq: Create quantum computing software without limits,''
  \url{https://www.classiq.io/}. [Online]. Available:
  \url{https://www.classiq.io/}
\BIBentrySTDinterwordspacing

\bibitem{bishop2017qasm}
L.~S. Bishop, ``Qasm 2.0: A quantum circuit intermediate representation,'' in
  \emph{APS March Meeting Abstracts}, vol. 2017, 2017, pp. P46--008.

\bibitem{dalcin2021mpi4py}
L.~Dalcin and Y.-L.~L. Fang, ``mpi4py: Status update after 12 years of
  development,'' \emph{Computing in Science \& Engineering}, vol.~23, no.~4,
  pp. 47--54, 2021.

\bibitem{haine2021middleware}
C.~Haine, U.-U. Haus, M.~Martinasso, D.~Pleiter, F.~Tessier, D.~Sarmany,
  S.~Smart, T.~Quintino, and A.~Tate, ``A middleware supporting data movement
  in complex and software-defined storage and memory architectures,'' in
  \emph{High Performance Computing: ISC High Performance Digital 2021
  International Workshops, Frankfurt am Main, Germany, June 24--July 2, 2021,
  Revised Selected Papers 36}.\hskip 1em plus 0.5em minus 0.4em\relax Springer,
  2021, pp. 346--357.

\bibitem{qiskit-aer}
J.~Doi and H.~Horii, ``Cache blocking technique to large scale quantum
  computing simulation on supercomputers,'' in \emph{2020 IEEE International
  Conference on Quantum Computing and Engineering (QCE)}.\hskip 1em plus 0.5em
  minus 0.4em\relax IEEE, 2020, pp. 212--222.

\bibitem{harris2020array}
\BIBentryALTinterwordspacing
C.~R. Harris, K.~J. Millman, S.~J. van~der Walt, R.~Gommers, P.~Virtanen,
  D.~Cournapeau, E.~Wieser, J.~Taylor, S.~Berg, N.~J. Smith, R.~Kern, M.~Picus,
  S.~Hoyer, M.~H. van Kerkwijk, M.~Brett, A.~Haldane, J.~F. del R{\'{i}}o,
  M.~Wiebe, P.~Peterson, P.~G{\'{e}}rard-Marchant, K.~Sheppard, T.~Reddy,
  W.~Weckesser, H.~Abbasi, C.~Gohlke, and T.~E. Oliphant, ``Array programming
  with {NumPy},'' \emph{Nature}, vol. 585, no. 7825, pp. 357--362, Sep 2020.
  [Online]. Available: \url{https://doi.org/10.1038/s41586-020-2649-2}
\BIBentrySTDinterwordspacing

\end{thebibliography}

\end{document}